\documentclass[pmlr]{jmlr}
\usepackage{mathtools}
\usepackage{multirow}
\usepackage{enumitem}
\usepackage{makecell}
\usepackage{longtable}

\usepackage{longtable}

\usepackage{booktabs}
\usepackage[load-configurations=version-1]{siunitx} 

\makeatletter
\def\set@curr@file#1{\def\@curr@file{#1}} 
\makeatother

\theorembodyfont{\upshape}
\theoremheaderfont{\scshape}
\theorempostheader{:}
\theoremsep{\newline}

\usepackage{algorithmic,algorithm}

\title[DrugChat: Towards Enabling ChatGPT-Like Capabilities on Drug Molecule Graphs]{DrugChat: Towards Enabling ChatGPT-Like Capabilities on Drug Molecule Graphs}

\author{\Name{Youwei Liang$^*$}
       \AND
\Name{Ruiyi Zhang$^*$}
      \AND
\Name{Li Zhang} 
\AND
\Name{Pengtao Xie}
       \AND
       \addr University of California San Diego,   *Equal contribution.
       }

\begin{document}

\maketitle

\begin{abstract}
A ChatGPT-like system for drug compound analysis could be a game-changer in pharmaceutical research, accelerating drug discovery, enhancing our understanding of structure-activity relationships, guiding lead optimization, aiding drug repurposing, reducing the failure rate, and streamlining clinical trials. In this work, we make an initial attempt towards enabling ChatGPT-like
capabilities on drug molecule graphs, by developing a prototype system DrugChat. DrugChat works in a similar way as ChatGPT. Users upload a compound molecule graph and ask various questions about this compound. DrugChat will answer these questions in a multi-turn, interactive manner. 
The DrugChat system consists of a graph neural network (GNN), a large language model (LLM), and an adaptor. The GNN takes a compound molecule graph as input and learns a representation for this graph. The adaptor transforms the graph representation produced by the GNN  into another  representation that is acceptable to the  LLM. The LLM takes the compound representation transformed by the adaptor and users' questions about this compound as inputs and generates answers. All these components are trained end-to-end.  To train DrugChat, we collected   instruction tuning datasets which contain 10,834 drug compounds and 143,517 question-answer pairs. The code and data is available at \url{https://github.com/UCSD-AI4H/drugchat}
\end{abstract}

\section{Introduction}

The process of drug discovery and development is a time-intensive and costly endeavor, often taking years and billions of dollars to bring a single drug to market~\citep{bill}. This process involves the exploration and understanding of vast chemical spaces and the intricate relationships between molecular structures and their biological activities, commonly known as structure-activity relationships (SAR)~\citep{sar}. Traditional methods~\citep{traditional}  often involve laborious iterative testing, with a high rate of late-stage failures. Recent advancements in computational chemistry and chemoinformatics~\citep{drug} have offered some respite, but there is still a pressing need for tools that can intuitively understand and generate meaningful insights from the complex data inherent in molecular graphs of drug compounds.

This technical report  introduces the concept of applying ChatGPT-like capabilities to drug molecule graphs, aiming to revolutionize the way we interact with and understand these complex entities. By transforming these molecular graphs into a form amenable to AI analysis, we can enable dynamic exploration of chemical spaces, efficient prediction of compound properties, and intelligent suggestions for drug design and optimization. A ChatGPT-like AI system capable of understanding drug compound molecule graphs and answering various questions about these drugs could revolutionize pharmaceutical research in several ways:
\begin{itemize}
    \item \textbf{Speeding up Drug Discovery}.  A ChatGPT-like AI system could drastically cut down the time required for initial stages of drug discovery by providing immediate insights into a compound's potential therapeutic uses, side effects, and contraindications based on its structure.
    \item \textbf{Predicting Drug Interactions}. A ChatGPT-like AI system could predict potential interactions between new drug candidates and existing drugs. By comparing the molecular structures of thousands of known substances, the system could identify possible conflicts or synergies, helping researchers to better anticipate how a new drug might behave in the real world.
    \item \textbf{Understanding Structure-Activity Relationships (SAR)}.  SAR~\citep{sar} is a crucial aspect of drug design. A ChatGPT-like AI system could help researchers understand the relationship between a drug's chemical structure and its biological activity. It could also help predict what modifications to the chemical structure might enhance its effectiveness or reduce unwanted side effects.
    \item \textbf{Guiding Lead Optimization}.  During the drug discovery process, `lead' compounds (those that show promise in initial screenings)~\citep{earlydrug} are typically optimized for better efficacy, reduced toxicity, and improved pharmacokinetics. A ChatGPT-like AI system could offer suggestions for structural modifications to enhance these parameters, guiding researchers in the right direction and saving valuable time.
    \item \textbf{Supporting Drug Repurposing}. A ChatGPT-like AI system could also aid in drug repurposing~\citep{repurpose} efforts. By understanding the structural properties of existing drugs, it could identify candidates that may be effective against diseases they were not initially developed to treat. This could help breathe new life into existing drugs and offer more rapid routes to treatment for challenging diseases. 
    \item \textbf{Reducing the Failure Rate}.   The failure rate in drug discovery is high~\citep{earlydrug}, often due to unforeseen toxicity or efficacy issues that emerge late in development. By providing more accurate predictions about a drug's properties and effects at the outset, a ChatGPT-like AI system could help reduce these costly late-stage failures.
    \item \textbf{Streamlining Clinical Trials}.  A ChatGPT-like AI could help design more effective clinical trials by predicting how a drug will interact with other medications or conditions, enabling researchers to target their trials more effectively and recruit suitable patient cohorts.
\end{itemize}


Developing a ChatGPT-like system for drug molecule graphs is highly challenging. First, one of the primary challenges lies in the representation of molecular graphs. Unlike text data, which is sequential and has a well-defined structure, molecular graphs are inherently non-sequential and highly complex, with no clear start or end point. Developing a methodology to translate these graphs into a format that a GPT model can process is crucial. Second, another significant challenge is capturing and understanding the intricate structure-activity relationships (SAR) in drug compounds. These relationships are often not straightforward and can involve subtle interactions between different parts of a molecule. Third, to train such a system, large datasets of molecular structures, along with their associated biological activities, side effects, and other properties, would be required. The generation or compilation of such datasets is a non-trivial task, given the proprietary nature of much of this information and the sheer diversity of the chemical space.

In this technical report, we make an initial attempt towards enabling ChatGPT-like
capabilities on drug molecule graphs, by developing a prototype system DrugChat. DrugChat works in a similar way as ChatGPT. Users upload a compound molecule graph and ask various questions about this compound. DrugChat will answer these questions in a multi-turn, interactive manner. 

The DrugChat system consists of a graph neural network (GNN)~\citep{pretraingnn}, a large language model (LLM)~\citep{vicuna}, and an adaptor. The GNN takes a compound molecule graph as input and learns a representation for this graph. The adaptor transforms the graph representation produced by the GNN  into another  representation that is acceptable to the  LLM. The LLM takes the compound representation transformed by the adaptor and users' questions about this compound as inputs and generates answers. All these components are trained end-to-end.  To train DrugChat, we collected   instruction tuning datasets which contain 10,834 drug compounds and 143,517 question-answer pairs.

The major contributions of this work are as follows: 
\begin{itemize}
\item We develop DrugChat, a prototype system aiming to achieve ChatGPT-like
capabilities on drug molecule graphs. DrugChat allows users to interactively ask open-ended questions about drug compounds and provides informative answers. 
\item We collect   instruction tuning datasets which contain 10,834 drug compounds and 143,517 question-answer pairs. These datasets enable the training of ChatGPT-like models for drug compounds and are publicly available. 
  \item To our best knowledge, DrugChat represents the first system that bridges graph data with large language models (LLMs), which enables interactive conversations on graphs. Our system seamlessly integrates graph neural networks with LLMs and can be easily extended to analyze other graph data beyond compound molecule graphs. 
\end{itemize}

\section{Related Works}
AI-based drug properties analysis~\citep{aidrug} is  a promising approach to significantly reduce costs and time associated with the traditional drug discovery and development pipeline. For example, ImageMol \citep{drug} is an unsupervised pretraining  method that processes images of chemical structures and outputs molecular properties, such as drug metabolism, brain penetration, toxicity, and molecular target profiles like beta-secretase enzyme and kinases. ImageMol was pretrained on 10 million unlabeled drug-like, bioactive molecules through predicting molecular targets of candidate compounds. It was assessed on 51 benchmark datasets and demonstrated high accuracy in identifying anti-SARS-CoV-2 molecules across 13 high-throughput experimental datasets.

Large language models (LLMs)~\citep{gpt3}  have demonstrated outstanding  capabilities in  generating inventive text, responding to reading comprehension queries, mathematical reasoning, etc. Nevertheless, the weight parameters of some of the most powerful LLMs are not publicly available, which considerably hinders academic research. Moreover, early LLMs were limited to processing text information as input, restricting their ability to understand information in other modalities. Consequently, researchers have developed an array of publicly accessible LLMs, including multi-modal variants, to address these challenges.

LLaMA~\citep{llama}, an LLM with 65 billion parameters, was developed by Meta AI \footnote{\url{https://ai.facebook.com/blog/large-language-model-llama-meta-ai/}}. This model is trained on publicly accessible datasets including English CommonCrawl, C4, Github, Wikipedia, Gutenberg Project, ArXiv, and Stack Exchange. This training dataset encompasses 20 languages with the most speakers. Although it is smaller than GPT-3, LLaMA demonstrates superior  performance on many  benchmarks, including  commonsense reasoning, question answering, reading comprehension, code generation, etc.  
Vicuna~\citep{vicuna} is an open-source chatbot trained by fine-tuning LLaMA on around 70,000 user-shared conversations with ChatGPT \footnote{\url{https://lmsys.org/blog/2023-03-30-vicuna/}}. 
 Vicuna was evaluated  using  GPT-4. 
MiniGPT-4~\citep{minigpt} is a vision-language model capable of processing images and generating textual answers  \footnote{\url{https://minigpt-4.github.io/}}. For instance, it can create stories and poems from input  images and offer solutions to problems depicted in images. MiniGPT-4 incorporates a  linear projection layer to align the visual information from a pretrained vision encoder with a  large
language model. 
LLaVA~\citep{llava} is another  multi-modal model that integrates  a vision encoder with an LLM  to facilitate general-purpose visual and language understanding \footnote{\url{https://llava-vl.github.io/}}. Different from MiniGPT-4, LLaVA is trained on  multi-turn conversation data. The projection layer and the LLM are both updated in the training stage while only the visual encoder parameters are frozen. The training data was created by converting raw image-text pairs into a suitable instruction-following format using ChatGPT/GPT-4. 

\section{Drug  Instruction Tuning Data}
To train DrugChat, we curate instruction tuning data for drug compounds, from two sources - ChEMBL and PubChem. Table~\ref{stat} summarizes the statistics of the curated two datasets. 

\begin{table}[ht]
\caption{Dataset statistics.}
\label{stat}
\begin{center}
\begin{small}
\begin{sc}
\begin{tabular}{l|c|c}
\toprule
Dataset&Number of Drugs&Number of Question-Answer Pairs\\
\midrule
ChEMBL&3,892&129,699\\
PubChem&6,942&13,818\\
\midrule
\end{tabular}
\end{sc}
\end{small}
\end{center}
\vskip -0.1in
\end{table}

\subsection{ChEMBL Drug Instruction Tuning Dataset}

In this section, we outline the process of constructing the ChEMBL Drug Instruction Tuning  dataset. We gathered data from the ChEMBL website\footnote{\url{https://www.ebi.ac.uk/chembl/}}, which houses information for a total of 2,354,965 chemical compounds. We  downloaded the sqlite version of the data dump, which was last updated on February 28, 2023\footnote{\url{https://ftp.ebi.ac.uk/pub/databases/chembl/ChEMBLdb/latest/}}. Out of the entire dataset, we identified 14,816 compounds that contain drug  information. After applying further filtering to exclude drugs with insufficient descriptive information, we ended up with a dataset containing 3,892 drugs.

For each drug, we initially gathered its SMILES string, representing the molecular structure. Subsequently, we obtained various molecular features, including the complete molecular formula and its classification as an acid, base, or neutral compound. Additionally, we collected drug-specific properties, such as the mechanism of action and  therapeutic applications. Based on these attributes, we manually crafted 129,699 question-answer pairs. Question-answer (QA) pairs for an exemplar drug in  the ChEMBL Drug Instruction Tuning Dataset is shown in Table 2. 

\begin{center}
    {Table 2. Question-answer (QA) pairs for an exemplar drug in  the ChEMBL Drug Instruction Tuning Dataset.}
\end{center}

\begin{small}
\begin{longtable}{l|c}
\toprule
SMILES of the drug &CC(=O)/N=c1/sc(S(N)(=O)=O)nn1C\\
\midrule
QA Pair 1& \makecell{\textbf{Q:} Please provide a description of this drug's mechanism of action.\\\textbf{A:} Carbonic anhydrase II inhibitor.} \\
\midrule
QA Pair 2& \makecell{\textbf{Q:} Is this compound an acid, a base, or neutral? \\ \textbf{A:} Neutral.} \\
\midrule
QA Pair 3& \makecell{\textbf{Q:} How many rotatable bonds does this compound have? \\ \textbf{A:} 1.} \\
\midrule
QA Pair 4& \makecell{\textbf{Q:} Determine if this drug is administered as a racemic  mixture, a single \\stereoisomer, an achiral molecule, or has an unknown chirality. \\ \textbf{A:} 
  An achiral molecule.} \\
\midrule
QA Pair 5& \makecell{\textbf{Q:} Does this compound satisfy the rule-of-three criteria? \\ \textbf{A:} No.} \\
\midrule
QA Pair 6& \makecell{\textbf{Q:} How many violations of Lipinski's Rule of Five are there for this \\compound, using the HBA\_LIPINSKI and  HBD\_LIPINSKI counts? \\ \textbf{A:} 0.} \\
\midrule
QA Pair 7& \makecell{\textbf{Q:} Is it known whether this drug is administered parenterally? \\ \textbf{A:} No.} \\
\midrule
QA Pair 8& \makecell{\textbf{Q:} Is this compound a small molecule polymer,\\ such as polystyrene sulfonate? \\ \textbf{A:} No.} \\
\midrule
QA Pair 9& \makecell{\textbf{Q:} What is the calculated ALogP value for this compound? \\ \textbf{A:} -1.42.} \\
\midrule
QA Pair 10& \makecell{\textbf{Q:} Is this molecule characterized by a small molecular\\ structure or a protein sequence?\\ \textbf{A:} It has a small molecule structure.} \\
\midrule
QA Pair 11& \makecell{\textbf{Q:} Determine if this molecule is inorganic, meaning it contains only metal\\ atoms and fewer than two carbon atoms. \\ \textbf{A:} No.} \\
\midrule
QA Pair 12& \makecell{\textbf{Q:} What is the highest development stage achieved for this compound \\across all indications? Please respond with Approved, Phase 3 Clinical\\ Trials, Phase 2 Clinical Trials, Phase 1 Clinical Trials, \\Early Phase 1 Clinical Trials, or Clinical Phase Unknown. \\ \textbf{A:} Approved.} \\
\midrule
QA Pair 13& \makecell{\textbf{Q:} What is the first recorded year of approval for this drug?  \\ \textbf{A:} 1959.} \\
\midrule
QA Pair 14& \makecell{\textbf{Q:} Is there a black box warning associated with this drug? \\ \textbf{A:} No.} \\
\midrule
QA Pair 15& \makecell{\textbf{Q:} Is it known if this drug is the first approved in its class, regardless of \\the indication or route of administration, acting on a specific target? \\ \textbf{A:} No.} \\
\midrule
QA Pair 16& \makecell{\textbf{Q:} How many hydrogen bond acceptors does this compound have? \\ \textbf{A:} 6.} \\
\midrule
QA Pair 17& \makecell{\textbf{Q:} Is this drug used for therapeutic purposes, rather than for imaging, \\additives, or other non-therapeutic applications? \\ \textbf{A:} Yes.} \\
\midrule
QA Pair 18& \makecell{\textbf{Q:} What is the polar surface area (PSA) value of this compound?\\ \textbf{A:} 107.41.} \\
\midrule
QA Pair 19& \makecell{\textbf{Q:} Has this approved drug been withdrawn due to toxicity\\ reasons for all indications,  populations, and \\doses in at least one country (not necessarily the US)? \\ \textbf{A:} No.} \\
\midrule
QA Pair 20& \makecell{\textbf{Q:} How many violations of Lipinski's Rule of Five (using HBA and\\ HBD definitions) are there for this compound? \\ \textbf{A:} 0.} \\
\midrule
QA Pair 21& \makecell{\textbf{Q:} How many heavy (non-hydrogen) atoms does this compound have? \\ \textbf{A:} 14.} \\
\midrule
QA Pair 22& \makecell{\textbf{Q:} How many aromatic rings does this compound have? \\ \textbf{A:} 1.} \\
\midrule
QA Pair 23& \makecell{\textbf{Q:} Determine the type of availability for this drug.\\ \textbf{A:} Prescription only.} \\
\midrule
QA Pair 24& \makecell{\textbf{Q:} Is it known whether this drug is taken orally?\\ \textbf{A:} Yes.} \\
\midrule
QA Pair 25& \makecell{\textbf{Q:} How many hydrogen bond acceptors are there in this compound, \\calculated according to Lipinski's original rules \\(i.e., counting N and O atoms)?\\ \textbf{A:} 7.} \\
\midrule
QA Pair 26& \makecell{\textbf{Q:} Is the drug administered\\ in this specific form, such as a particular salt?\\ \textbf{A:} Yes.} \\
\midrule
QA Pair 27& \makecell{\textbf{Q:} Determine if this compound is a prodrug.\\ \textbf{A:} No.} \\
\midrule
QA Pair 28& \makecell{\textbf{Q:} What is the molecular weight of this compound's parent molecule?\\ \textbf{A:} 236.28.} \\
\midrule
QA Pair 29& \makecell{\textbf{Q:} How many hydrogen bond donors are there in this compound, \\calculated according to Lipinski's original\\ rules (i.e., counting NH and OH groups)?\\ \textbf{A:} 2.} \\
\midrule
QA Pair 30& \makecell{\textbf{Q:} Which USAN substem can\\ this drug or clinical candidate name be matched with?\\ \textbf{A:} Zolamide.} \\
\midrule
QA Pair 31& \makecell{\textbf{Q:} What is the definition of this compound's USAN stem?\\ \textbf{A:} Carbonic anhydrase inhibitors.} \\
\midrule
QA Pair 32& \makecell{\textbf{Q:} What is the molecular formula\\ of this compound, including any salt that it may have?\\ \textbf{A:} C5H8N4O3S2.} \\
\midrule
QA Pair 33& \makecell{\textbf{Q:} How many hydrogen bond donors does this compound have?\\ \textbf{A:} 1.} \\
\midrule
QA Pair 34& \makecell{\textbf{Q:} What is the classification of this molecule? Please respond with \\Small Molecule, Protein, Antibody, Oligosaccharide, Oligonucleotide,\\ Cell, Enzyme, Gene, or Unknown.\\ \textbf{A:} Small molecule.} \\

\bottomrule

\end{longtable}
\vskip -0.1in
\end{small}

\subsection{PubChem  Drug Instruction Tuning Dataset}

In this section, we outline the process of constructing the PubChem  Drug Instruction Tuning Dataset. The data was collected from the PubChem website\footnote{\url{https://pubchem.ncbi.nlm.nih.gov/}}, which houses information of  66,469,244 chemical compounds. We utilized the data version last updated on May 9, 2023\footnote{\url{https://pubchem.ncbi.nlm.nih.gov/classification/}}. Among these compounds, 19,319 possess drug information. After filtering out drugs lacking detailed textual descriptions, we retained 6,942 drugs.

For each drug, we gathered its SMILES string and descriptions from databases including ChEBI, LOTUS, and YMDB. In total, we acquired 13,818 compound description summaries and curated 13,818 question-answer (QA) pairs. QA pairs for an exemplar drug in PubChem are  provided in Table 3.


\begin{table}[ht]

\caption{Question-answer (QA) pairs for an exemplar drug in  the PubChem Drug Instruction Tuning Dataset.} 
\label{exp2}
\begin{center}

\begin{small}
\begin{tabular}{l|c}
\toprule
Compound Name&Acetylcarnitine\\
\midrule
Canonical SMILES&CC(=O)OC(CC(=O)[O-])C[N+](C)(C)C\\
\midrule
ChEBI Database& \makecell{
\textbf{Q:} Please describe this drug.\\\textbf{A:} O-acetylcarnitine is an O-acylcarnitine \\having acetyl as the acyl substituent. \\It has a role as a human metabolite. \\It is functionally related to an acetic acid. \\It is a conjugate base of an O-acetylcarnitinium.} \\
\midrule
LOTUS Database& \makecell{\textbf{Q:} Please describe this drug.\\\textbf{A:} Acetylcarnitine is a natural \\product found in Pseudo-nitzschia multistriata, \\Euglena gracilis, and other organisms with data available.} \\
\midrule
YMDB Database& \makecell{\textbf{Q:} Please describe this drug.\\\textbf{A:} L-Acetylcarnitine is a metabolite \\found in or produced by Saccharomyces cerevisiae.} \\
\midrule
\end{tabular}
\end{small}
\end{center}
\vskip -0.1in
\end{table}

\section{Method}
\begin{figure}
    \centering
    \includegraphics[width=\textwidth]{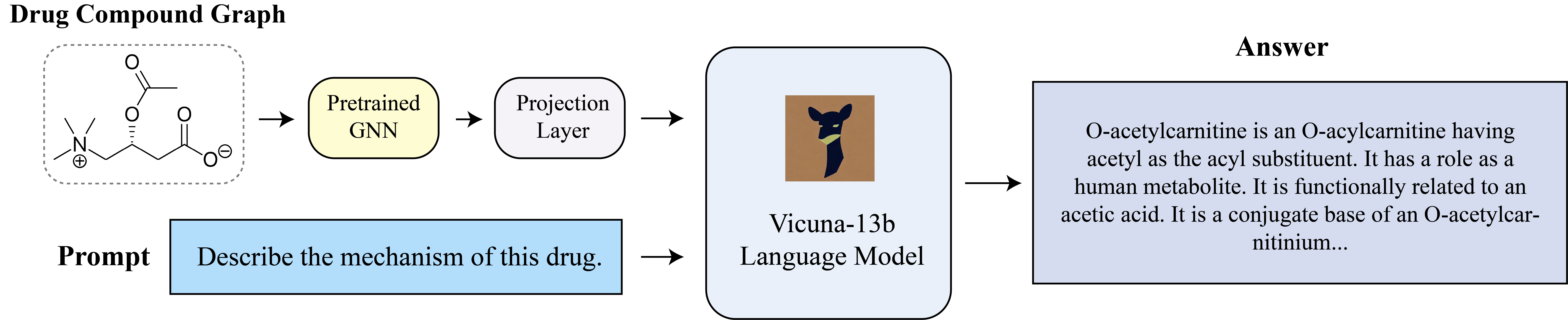}
    \caption{An overview of the DrugChat framework.}
    \label{fig:1}
\end{figure}

An overview of DrugChat is provided in Figure \ref{fig:1}.
It takes a compound molecule graph as input and allows users to ask multi-turn questions about this compound. For each question, DrugChat generates an answer. DrugChat consists of a graph neural network (GNN), a large language model (LLM), and an adaptor between GNN and LLM. The GNN learns a  representation for the  compound molecule graph. The adaptor (which is a linear transformation matrix) transforms the graph representation into an LLM-compatible soft prompt vector.  The LLM takes a user-question and the graph  prompt vector as inputs and generates an answer. 
We employ a pretrained GNN from  \citep{pretraingnn} and a pretrained LLM - Vicuna13b~\citep{vicuna}. When training DrugChat, we fix the weight parameters of the GNN and LLM, and only update the adaptor's weights. Given a drug in the  instruction tuning data and a question about this drug, the drug's graph is first fed into the GNN to produce a representation vector which is then fed into the adaptor to produce a prompt vector. The prompt vector and the question are fed into the LLM to generate an answer. A negative log likelihood loss  between the generated answer and groundtruth answer is calculated. The adaptor is trained by minimizing this loss. Next, we introduce each component in DrugChat. 


\begin{figure}
    \centering
    \includegraphics[width=0.9\textwidth]{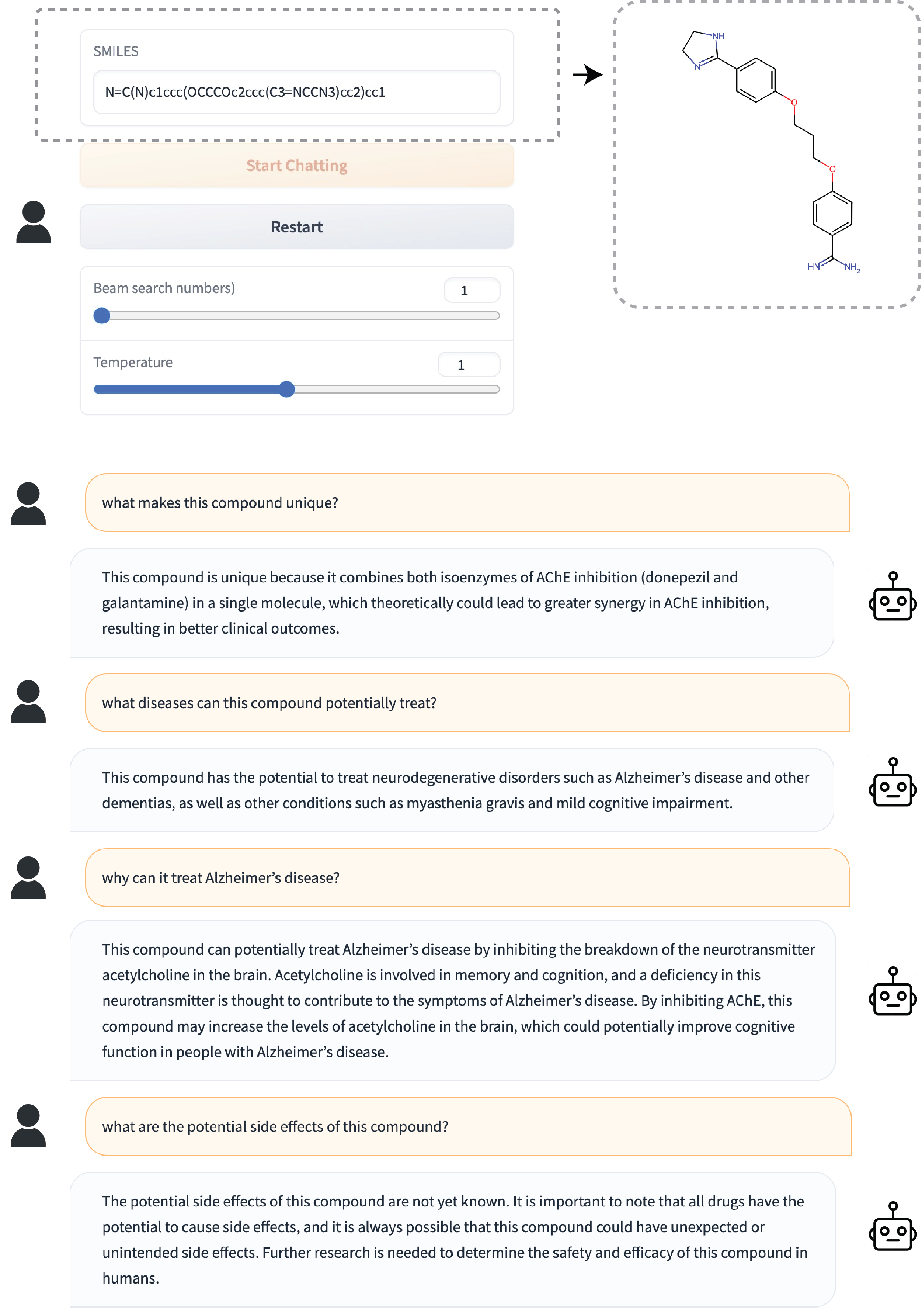}
    \caption{A demonstration of the DrugChat framework.}
    \label{fig:2}
\end{figure}

\begin{figure}
    \centering
    \includegraphics[width=\textwidth]{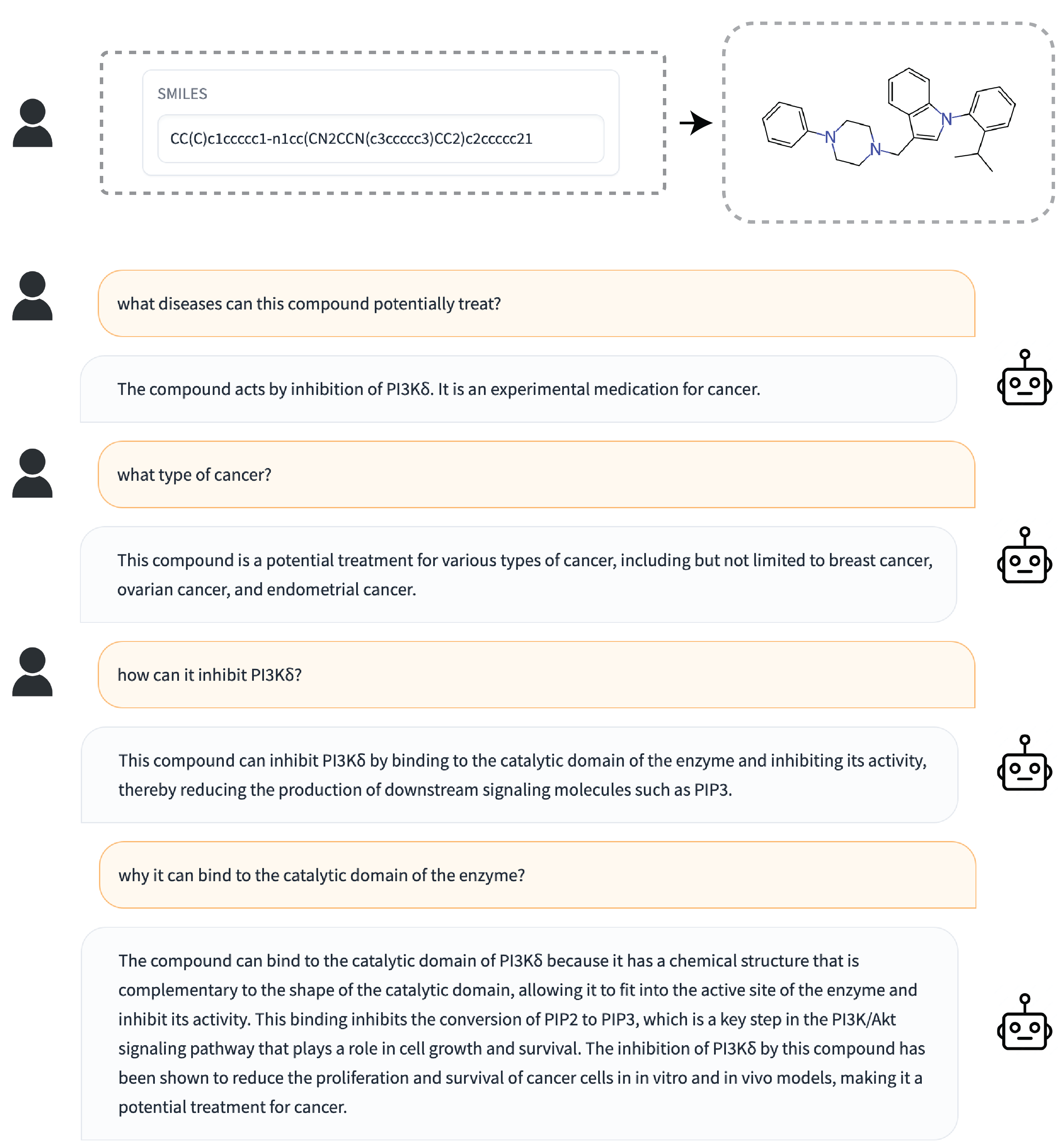}
    \caption{A demonstration of the DrugChat framework.}
    \label{fig:3}
\end{figure}

\subsection{Graph Neural Networks (GNNs)} A GNN~\citep{pretraingnn}  utilizes the graph structure and node features to learn multiple layers of  representation vectors  for each node $v$ and a representation vector $h_G$ for the entire graph $G$. The representation of each node is  updated by aggregating features from its neighboring nodes.   
At layer $k$, the  representation $h_v^k$ of node $v$ encompasses information from nodes within the $k$-hop network neighborhood of $v$.  Formally, $h_v^k$ is calculated as follows:
\begin{equation}
    h_v^k=\sigma(h_v^{k-1},AGG(\{h_u^{k-1},u\in \mathcal{N}(v)\})),
\end{equation}
where $AGG$ represents a function that aggregates information from neighboring nodes. $\sigma$ is a function that combines information from the node feature of the previous layer and  neighboring nodes. $\mathcal{N}(v)$ indicates all neighboring nodes for node $v$. To acquire the representation vector for the entire graph $G$, a pooling function $f$ (which is permutation-invariant, such as averaging) extracts information from all node features at the final layer $K$:
\begin{equation}
    h_G=f(\{h_v^K,v\in G\}). 
\end{equation}

\subsection{Large Language Models (LLMs)} 

LLMs  utilize the Transformer~\citep{transformer} decoder to model the conditional  probability $p_\theta(n_i|n_{<i})$ for token $n_i$ 
in a  language model. The Transformer decoder applies a multi-head self-attention block~\citep{transformer} to the input context tokens and uses a position-wise feedforward network to calculate the  probabilities over output tokens. Given the context vector of tokens, the token generation probability is computed as follows:

\begin{equation}
\begin{split}
    &h_0=N_{i-1}W_e+W_p \\
    &h_l=transformer\_block(h_{l-1})(1\leq l\leq m) \\
    &p(n_i|N_{i-1})=Softmax(h_mW_e^T)
\end{split}
\end{equation}
where $N_{i-1}=(n_1,n_2,...,n_{i-1})$ denotes the context vector of tokens, $m$ refers to the number of layers, $W_e$ denotes the token embedding matrix, and $W_p$ represents the positional embedding matrix.

\subsection{Aligned Graph-Text Generation}

At this stage, we create a prompt for every training graph-text pair, which allows the LLM to generate descriptions from drug compound graphs. We utilize a prompt template that adheres to the conversational format of Vicuna-13b:

\begin{equation}
\begin{split}
    &\textbf{Q:} <Graph><GraphFeature></Graph><Instruction> \\
    &\textbf{A:} <Desc>\\
\end{split}
\end{equation}
    
In this prompt, $<GraphFeature>$ is a soft prompt that symbolizes the graph structure feature encoded by the linear projection layer. $<Instruction>$ serves as a directive sentence, prompting the LLM to generate descriptions for the drug, such as ``Describe the mechanism of this drug''.  During the training stage, $<Desc>$ is populated with descriptive text from human experts to train the linear projection layer. In the testing stage, $<Desc>$ remains empty, and the model is expected to generate descriptive text for the provided drug structure.

\section{Results}
We tested DrugChat on compound graphs that are not contained in the training data. Figure 2 and 3 show two examples. DrugChat can answer diverse multi-turn questions about compounds, such as ``what makes this compound unique?'', ``what diseases can this compound potentially treat?''. Note that these questions are not contained in the training data. We will perform a systematic quantitative evaluation by collaborating with pharmaceutical scientists.  


\section{Conclusions,  Limitations and Future Work}

In this report, we present the DrugChat framework, designed to answer questions and generate text descriptions for drugs from their molecular graphs. We develop the ChEMBL Drug  Dataset and the PubChem Drug  Dataset to train the DrugChat model. With further development and evaluation,  DrugChat has the potential to enable conversational analysis of drug compounds.

A potential limitation of DrugChat is language hallucination. Since DrugChat incorporates an LLM module, it may occasionally produce untrustworthy answers and descriptions for drugs, which hampers its practical application in real drug discovery pipelines. If DrugChat generates seemingly convincing yet incorrect text descriptions, it could mislead human decision-makers and potentially result in adverse consequences.

In the future, this issue could be mitigated by utilizing higher-quality training data and implementing effective filtering strategies. Additionally, more advanced GNN encoders and LLMs will play a crucial role in addressing this challenge. As the number of users increases, human feedback can also be utilized to fine-tune the DrugChat model through reinforcement learning techniques \citep{instructgpt}.

\bibliography{neurips}

\end{document}